\documentclass[12pt]{article}
\usepackage{a4wide}
\usepackage{amsmath, amssymb, epsf, euscript}
\usepackage{nrj,pseudocode}

\title{Energy Optimal Data Propagation in Wireless Sensor Networks}
\author{Pierre Leone, Olivier Powell, Jos\'e Rolim
\thanks{\{pierre.leone$|$olivier.powell$|$jose.rolim\}@cui.unige.ch}}
\begin{document}
%\setcounter{page}{0}
%\tableofcontents
\maketitle

\begin{abstract}
We propose an algorithm which produces a randomized strategy reaching optimal
data propagation in wireless sensor networks (WSN).
In \cite{nrj:balance:jose} and \cite{nrj:balance:pierre}, an energy balanced solution is sought
using an approximation algorithm. Our algorithm improves by 
(a) when an energy-balanced solution does not exist, it still finds an optimal solution 
(whereas previous algorithms did not consider this case and provide no useful solution)
(b) instead of being an approximation algorithm, it finds the exact solution in one pass.
We also provide a rigorous proof of the optimality of our solution.
\end{abstract}
%\keywords{sensor networks, blablabla}

\section{Introduction}
In \cite{nrj:balance:jose}, the problem of finding and energy-balanced solution to data propagation
in a Wireless Sensor Network (WSN) is considered. They model a WSN by splitting in it in slices,
an try to maximize lifespan of the network by ensuring that the expected energy consumption in each slice
is the same. The case they consider is the following: sensors are distributed randomly and informally in a
circle (or a sector), and data has to be propagated from the WSN towards the sink, which is in the center of the circle.
It is considered that data is to be slided towards the sink 
along the network in  a multi-hop fashion, but that any node which contributes
to this propagation can decide to eject the message directly to the sink in an energy expensive single-hop long range data transmission.
Using this model, the authors of
\cite{nrj:balance:jose} show that energy-balance can be reached if a recurrence relation between the 
\emph{probabilities that a slice ejects a message single-hop to the sink} is satisfied.
They then propose an approximation algorithm (converging exponentially fast towards the solution) to compute the solution
to the energy-balancing problem.
Although they do not prove it, it happens to be that this energy-solution always exists in the context they have chosen: a
circle (or a sector) with informally distributed sensor nodes inside of it, and a uniform distribution of events 
(which generate data to be slided towards the sink) inside the WSN.
In \cite{nrj:balance:pierre}, a more general case is studied: events may not happen according to a uniform distribution,
and the sensors may not be distributed uniformly over the area to be monitored, Moreover it is not assumed that
the distribution of events is known \emph{a priori}. A solution to the energy-balance problem is then computed on-line 
(by the central sink) whilst the distribution is discovered from observation. They show that their algorithm tends
to the energy-balanced solution (when this solution exists). 
However, neither \cite{nrj:balance:jose} nor \cite{nrj:balance:pierre} consider cases where an energy-balanced solution
does not exist. 
The idea we want to fulfill is to find an optimal solution, whatever the distribution of events and the
distribution of sensors is. We provide an algorithm to compute such a solution, which turns out to be the energy-balanced
solution when this solution exists. Another improvement is that our algorithm, in the contrary to the algorithm 
provides an exact optimal solution,
whereas the algorithms from \cite{nrj:balance:jose} and \cite{nrj:balance:pierre} are approximation
algorithms converging exponentially, but with potential infinite run time.
Our algorithm is based on a very simple idea, 
which can informally be stated as \emph{if your neighbour has more energy available then you, let him do the job}.
Although this is over-simplifying and special cases have to be treated for the general solution,
we propose as an application of this idea
a \emph{completely distributed} solution (i.e. not computed in a centralized sink) 
which converges to the optimal-solution under some reasonable assumptions,
and provide some simulation evidence of the interest of this approach.
\section{Model}
\label{section:model}
The model we study is the same as the one from \cite{nrj:balance:jose}
and \cite{nrj:balance:pierre}, and is the
following.
A \emph{sensor network} is composed of $ N $ \emph{slices} $ \set{ S_i }_{ 1 \leq i
  \leq N} $. Each of these slices represents a subset of the sensors
of the network.
For $ 2 \leq j \leq N - 1 $, the  $ j $th slice, $ S_j $, has two neighbour slices:
\begin{enumerate}
  \item $ S_{ j + 1 } $, which is called the \emph{previous slice}
  (with respect to $S_i$).
  \item $ S_{ j - 1 } $, which is called the \emph{next slice}.
\end{enumerate}
The first slice, $ S_1 $, is also neighbour to a special node of the network,
called the \emph{sink}. The last slice, $S_N $ has no previous slice,
but only a next slice: $ S_{ N - 1 } $. 

\begin{definition}
We say that a slice \emph{detects} an event if one of the sensors of
the slice detects an event.
\end{definition}
When a slice $S_i$ detects an event, it needs to report by sending a message
to the sink.
\begin{definition}
  A \emph{generated message} is a message which has to be sent by a slice $ S_i $ following
  the detection of an event by $ S_i $.
 \end{definition}
When a slice is in presence of a \emph{generated message}, it has the choice between two strategies:
\begin{enumerate}
  \item Send the message directly to the sink: we say that the (generated) message is \emph{ejected} by $ S_i $.
  \item Send the message to the next slice, $S_{ i - 1 }$, putting the responsibility
  of communicating the message to the sink on $S_{ i - 1 }$: we say that the
  (generated) message is \emph{slided} along the network from $ S_i $ towards $ S_{ i - 1 }$.
\end{enumerate}
\begin{definition}
  Once a generated message has been slided along the network,
  we do not call it a generated message anymore, but we rather call it a \emph{sliding message}.
\end{definition}
When a slice receives a \emph{sliding} message from its
previous slice, it has the choice between the two same strategies as for generated messages:
%%%%%%%%%%%%%%%%%%%%%%%%%%%%%%%%%%%%%%%%%%%%%%%%%%%%%%%%%%%%%%%%%%%%%%%%%%%%%%%%%%%%%%%%%%%%%%%%%%%%%%%
%%This is not as nice, but it takes less room
  either send the message to its next slice (the message keeps
  sliding along the network towards the sink) or
  stop the message, and send it directly to the sink: we say
  that the \emph{sliding} message is \emph{ejected} to the sink.
%% This is nice, but it takes a lot of room
%% \begin{enumerate}
%% \item 
%%   Send the message to its next slice (the message keeps
%%   sliding along the network towards the sink).
%% \item 
%%     Stop the message, and send it directly to the sink: we say
%%     that the \emph{sliding} message is \emph{ejected} to the sink.
%% \end{enumerate}
%%%%%%%%%%%%%%%%%%%%%%%%%%%%%%%%%%%%%%%%%%%%%%%%%%%%%%%%%%%%%%%%%%%%%%%%%%%%%%%%%%%%%%%%%%%%%%%%%%%%%%%
With this model in mind, we describe a slice and its behaviour by the following parameters.
\begin{itemize}
\item 
  $ g_i $ is the number of events being detected by $ S_i $.
\item 
  $ b_i $ is the amount of energy available in slice $S_i$,
  which is proportional to the number of sensors, assuming
  each sensor has the same amount of energy available.
\item 
  $ d_i $ is the distance of $ S_i $ to the sink.\footnote{The
  distance of a slice to 
  the sink is the maximum 
  over all distances between sensors of the slice and the sink. 
  We suppose w.l.o.g. that $ d_1 \geq 1 $ for every $ i $, 
  and  that $ d_{ i + 1 } \geq d_i $.}
\item 
  $ f_i $ is the number of messages slided from $ S_i $ towards $ S_{ i - 1 } $.
\item 
  $ j_i $ is the number of messages ejected from $ S_i $ to the sink.
\item
  $ p_i $ is the sliding probability: $ p_i = \frac{ f_i }{ f_i + g_i } $.
\item 
  $ \epsilon_i $ is the \emph{ejection probability} by slice $ S_i $ of a \emph{sliding} message.\footnote{
  That is, $ \epsilon_i $ is the fraction of the $ f_{ i + 1} $ sliding messages coming from
  $ S_{ i+1 } $ which are ejected from $S_i$ to the sink. E.g. if $
  g_i = 0 $, then $ \epsilon_i = 1 - p_i $.
}
\end{itemize}
Notice that the following recurrence relation holds
\begin{equation}
  f_i + j_i = f_{ i + 1 } + g_i 
\end{equation}
Transmitting messages has a price, which has to be payed by the transmitter.
In this model, we take the following convention.\footnote{
With this convention, we do not consider the cost of receiving a message.
However, this does not yield much loss of generality.
For example, if we wanted to fix the cost of receiving a message to be equal to
short range transmission (i.e. $1$ energy unit), we could include this in the model
by setting the price of long range transmission to $d_i^2+1$ and the price of short range transmission
to $2$.
}
\begin{convention}
Sliding a message from $ S_i $ towards $ S_{ i - 1 } $ costs $ 1 $ energy unit to $ S_i $,
and ejecting a message from $ S_i $ to the sink costs $ d_i^2 $ energy units to $ S_i $. 
\end{convention}
Therefore, the mean energy consumption of slice $ S_i $ is equal to the following:
\begin{equation}
  E_i := f_i + j_i d_i^2 
\end{equation}
\begin{definition}
  We say that the network is energy balanced if for $ 2 \leq N $ it holds that
  \begin{equation}
    \frac{ 1 }{ b_{ i } } E_{ i } = \frac{ 1 }{ b_{ i - 1 } } E_{ i - 1 }
  \end{equation}
\end{definition}
That is, the network is energy balanced if the mean energy consumption is the same for every \emph{sensor} of the network.
We next want to introduce a measure of the lifespan of a network 
pursuing the idea that the network is down as soon as
any single slice runs out of battery. The lifespan of the network thus depends
on the worst case energy consumption amongst slices.
\begin{definition}
  \label{df:lifespan}
  The lifespan of the network is defined as $ \min\set{ \frac{ b_i }{ E_i } }_{ 1 \leq i \leq N }$.
\end{definition}
Let $ m $ be a message either generated in or slided to the slice $ S_i $. 
Suppose $ S_i $ ejects the message directly to the sink with probability $ q $.
The mean energy consumption per sensor of slice $ S_i$ induced by the treatment of message $ m $ is equal to
\begin{equation}
  \frac{1}{b_i}\left[\left( 1 - q \right) + q d_i^2\right]
\end{equation}
Since message $ m $ is slided towards $ S_{ i - 1 } $ with probability $ 1- q $, it generates an average
energy consumption per sensor of $S_{ i - 1 } $ equal to the following:
\begin{equation}
  \label{cost:nextSlice}
  \frac{1}{b_{i-1}}\left( 1 - q \right)\left[\left( 1 - \epsilon_{ i - 1 } \right) + \epsilon_{ i - 1 } d_{ i - 1 }^2 \right]
\end{equation}
since $ \epsilon_{ i - 1 } $ is the ejection probability of a sliding message for $ S_{ i - 1 } $.
Therefore, we get the following claim.
\begin{claim}
  \label{epsilon:implication}
  Let $ m $ be a message that has to be treated by $ S_i $.
  Suppose that $ m $ is slided (by $ S_i $) along the network towards $ S_{ i - 1} $ with probability $ 1 - q $ 
  and that for for all $ 2 \leq i < N $ the following recurrence relation holds:
  \begin{equation}
    \label{rec:epsilon:gross}
    \frac{1}{b_{ i } }\left[\left( 1 - \epsilon_{ i } \right) + \epsilon_ i d_i^2\right] =
    \left( 1 - \epsilon_{ i } \right)\frac{1}{b_{ i - 1 } }\left[\left( 1 - \epsilon_{ i - 1 }\right) + \epsilon_{ i - 1 } d_{ i - 1 }^2 \right]
  \end{equation}
  Then if $ q = \epsilon_i $
  the treatment of message $ m $ (which implies contribution by all slices $ S_i $, $ S_{ i - 1 } $, $ \ldots $, $ S_1 $) 
  generates the same mean energy consumption per sensor in all slices $ S_i, \ldots, S_1 $. 
  \end{claim}
  This follows from the fact that treating $ m $ has an average cost per sensor of slice $ S_i $ (when $ q = \epsilon_i $)
  of $ m = \frac{ 1 }{ b_i }\left( 1 - \epsilon_i + \epsilon_i d_i^2\right) $.
  The mean average cost per sensor of $ S_{ i - 1 } $ is then seen to be equal to $ m $ too, because 
  of (\ref{rec:epsilon:gross}) and (\ref{cost:nextSlice}), and so on for $ S_{ i - 2 } $ to $ S_1 $ (details left to the reader). 
  
  Relation (\ref{rec:epsilon:gross}) can be rewritten in the following form:
\begin{equation}
  \label{rec:epsilon:fine}
  \epsilon_{ i + 1 } = \frac{ \frac{ b_{ i + 1 } }{ b_i } \left( 1 + \epsilon_{ i - 1 } { d_i }^2 - 1 \right) - 1 }{ \frac{ b_{ i + 1 } }{ b_i } \left( 1 + \epsilon_{ i - 1 } { d_i }^2 - 1 \right) + { d_{ i + 1 } }^2 - 1 }
\end{equation}
Setting $ \epsilon_1 = 1 $ (which is natural since the first slice can only send messages directly to the sink), we can compute directly the $ \epsilon_i $'s satisfying the recurrence relation for fixed $ b_i $'s and $ d_i $'s.
We next show how to use this fact to find an optimal solution for the
problem of lifespan maximization.

\section{The Algorithm}
\label{algorithm}
The algorithm we propose is an algorithm which computes a randomized strategy for data propagation in 
a sensor network separated in $N$ slices of nodes at increasing distances from the sink.
The \emph{input} is a description of the network and a statistical description of the data to be propagated
in the form of three sequences of same length 
$\set{b_i}_{1\leq i \leq N}$, $\set{d_i}_{1\leq i \leq N}$ and $\set{g_i}_{1\leq i \leq N}$
where the $b_i$'s describe the energy available in each slice, the $d_is$'s are the distances of each slice to the sink,
and the $ g_i $'s are the distribution of events generating data to be propagated in the network (c.f. section \ref{section:model}).
The \emph{output} is a sequence $\set{p_i}_{1\leq i \leq N}$ representing a randomized strategy the network should apply
in order to maximize its lifespan: each slice $S_i$ for $1\leq i\leq N$, when treating (either a sliding or a generated) message should
send the data directly to the sink with probability $ 1-p_i $ and slide it along the network to the next slice with probability $p_i$.
We can now start the description of the algorithm

Since the $ b_i $'s and $ d_i $'s are given from the input for $ 1 \leq i \leq N $,
we can compute the $ \epsilon_i $'s ( $ 1 \leq i \leq N $) which are the solution to relation (\ref{rec:epsilon:gross}) with $ \epsilon_1 = 1 $.
\begin{remark}
  \label{positive:epsilons}
  The attentive reader may have noticed that although it holds that equation ($\ref{rec:epsilon:fine}$) implies that
  $ \epsilon_i \leq 1 $, nothing guarantees that $ \epsilon_i \geq 0 $.
  For simplicity, let us make the temporary assumption that $ \epsilon_i \geq 0 $.
  We treat the case with negative $ \epsilon $'s in section \ref{second:problem}.
\end{remark}
For each $ 1 \leq i \leq N $, we define the following:
\begin{enumerate}
  \item $ G_i $ is the number of messages to be treated at slice $ S_i $.
  \item $ F_i $ is the number of messages forwarded from slice $ S_{ i+1 } $ towards slice $ S_{ i } $.
  \item $ J_i $ is the number of messages ejected from slice $ S_i $ directly to the sink.
  \item $ E_i $ is the average energy spent by slice $ S_i $, which is equal to $ F_i  + J_i { d_i }^2 $.
\end{enumerate}
Using the $ g_i $'s from the input
we initialize $ F_i = J_i = 0 $ and $ G_i = g_i $ for every $ 1 \leq i \leq N $.
We then treat each slice one at a time, in a top-down fashion (i.e. from $ S_1$, close to the sink, towards $ S_N $, away from the sink ):
\begin{itemize}
  \item Let $ J_1 = G_1 $.
    That is, all the messages generated at $ S_1 $ are ejected to the sink.
    This means an average energy consumption in slice $ S_1 $ of $ E_1
    = G_1 d_1^2 = J_1 d_1^2 + F_1 $.
  \item To take account of the fact that all the $G_1$ messages to be
  treated have been ejected, we update the value of $G_1$ to $G_1 = 0$.
  
\end{itemize}
We then repeat the following for each of the slices $ S_i $ for $ i $ from $ 2 $ to $ N $:
First, let $ J_i := \frac{ b_i }{ b_{ i - 1 } { d_i }^2 } E_{ i - 1 } $. 
This means that the following holds:
\begin{equation}
  \label{nrj:balance:firstStep}
  \frac{ 1 }{ b_i }J_i { d_i }^2 = \frac{ 1 }{ b_{ i - 1 } }E_{ i - 1 }
\end{equation}
The interpretation of equation (\ref{nrj:balance:firstStep}) is that the average cost per sensor in slice $ S_i $ due to ejection of messages is equal to the average energy consumption per sensor (so far) in slice $ S_{ i - 1 } $. 
So at this points, the average energy consumption per sensor in slice
$ S_i $ is equal to the average consumption per sensor in slice $ S_{
  i - 1 } $ (and, by induction, in $ S_{ i - 2 }, S_{ i - 3 }, \ldots$ ).
Notice that if $ J_i > G_i $, it means we are trying to eject more
messages than the total amount of messages available to be treated,
which is not physically possible.
Therefore, for our approach to work smoothly it should be that there
are enough messages to
be treated at slice $ S_i $, i.e. it should be that the following holds:
\begin{equation} 
  \label{cond:1}
  J_i = \frac{ b_i }{ b_{ i - 1 } { d_i }^2 } E_{ i - 1 } \leq G_i
\end{equation} 
This is an important condition, which does not hold without loss of generality.
We explain how to overcome this limitation in section \ref{first:problem}.
For the time being, we consider only the case where the initial $ G_i $'s are large enough so as to ensure that equation (\ref{cond:1}) holds.
Although so far energy balance is reached for the slices $ S_i $ to $ S_1 $ (because of equation (\ref{nrj:balance:firstStep}) and by induction),  
there are still $ G_i - J_i $ messages generated at slice $ S_i $ which have to be treated.
Let us update $ G_i $ to $ G_i := G_i - J_i $, thus $ G_i $ now counts the number of messages generated at slice $ G_i $ yet to be treated.
These messages have to be treated in such a way that they will increase the average energy spent per sensor in each of the slices $ S_i $ to $ S_1 $ by exactly the same amount. 
The strategy is the following: a fraction $ \left( 1 - \epsilon_i \right) $ of the $ G_i $ messages yet to be treated are forwarded to the next slice $ S_{ i - 1 } $, while the rest is ejected directly to the sink, thus increasing the average energy spent per sensor at slice $ S_i $.
Formally, this means setting the following:
\begin{eqnarray} 
  F_i & := & \left( 1 - \epsilon_i \right) G_i \label{first:of:many}\\
  J_i & := & J_i + \epsilon_i G_i
\end{eqnarray}
Of this fraction $ G_i \left( 1 - \epsilon_i \right) $ of messages slided from $S_i$ towards the slice $ S_{ i - 1 } $, a fraction $ 1 - \epsilon_{ i - 1 } $ will be further slided from $ S_{ i - 1 } $ towards $ S_{ i - 2 } $, while the rest is ejected directly to the sink from $ S_{ i - 1 } $, etc...
Algorithmically, this means doing the following:\\
\begin{eqnarray} 
  F_{ i - 1 } & := & F_{ i - 1 } + \left( 1 - \epsilon_{ i - 1 } \right) \left( 1 - \epsilon_{ i } \right) G_i \\
  J_{ i - 1 } & := & J_{ i - 1 } + \epsilon_{ i - 1 } \left( 1 - \epsilon_{ i } \right) G_i \\
  F_{ i - 2 } & := & F_{ i - 2 } + \left( 1 - \epsilon_{ i - 2 } \right) \left( 1 - \epsilon_{ i - 1 } \right) \left( 1 - \epsilon_{ i } \right) G_i \\
  J_{ i - 2 } & := & J_{ i - 2 } + \epsilon_{ i - 2 } \left( 1 - \epsilon_{ i - 1 } \right) \left( 1 - \epsilon_{ i } \right) G_i \\
  F_{ i - 3 } & := & F_{ i - 3 } + \left( 1 - \epsilon_{ i - 3 } \right) \left( 1 - \epsilon_{ i - 2 } \right) \left( 1 - \epsilon_{ i - 1 } \right) \left( 1 - \epsilon_{ i } \right) G_i \\
  J_{ i - 3 } & := & J_{ i - 3 } + \epsilon_{ i - 3 } \left( 1 - \epsilon_{ i - 2 } \right) \left( 1 - \epsilon_{ i - 1 } \right) \left( 1 - \epsilon_{ i } \right) G_i \\
  \label{last:of:many}
  \ldots & &
\end{eqnarray}
During this phase\footnote{Also, after this we can set $ G_i = 0 $,
  since all messages at slice $S_i$ have been treated.} (forwarding/ejecting from slice to slice), the energy consumption per sensor at slice $ S_i $ has been increased by 
$ m_i := \left( 1 / b_i \right) G_i \left[ \left( 1 - \epsilon_i \right) + \epsilon_i d_i ^2 \right] $, 
the increase of energy spent per sensor at slice $ S_{ i - 1 } $ is equal to 
$ m_{ i - 1 } := \left( 1 / b_{ i - 1 }\right)G_i \left( 1 - \epsilon_i \right)\left[ \left( 1 - \epsilon_{ i - 1 } + \epsilon_{ i - 1 } d_{ i - 1 }^2 \right) \right] $, etc... 
But because the $ \epsilon_i $'s satisfy equation (\ref{rec:epsilon:fine}), all these $m_j$'s ($ 1 \leq j \leq i $) have the same value, as follows from claim \ref{epsilon:implication}. 
So when we finish treating slice $ S_i $ the average energy consumption per sensor in $ S_1, S_2, \ldots, S_i $ is equal 
(slightly increased from the previous step), etc...
We then go on to treating the next slice ($S_{i+1}$) until we reach the last slice, $ S_N $. 
At this stage, the average energy consumption per sensor in each of
the slices will be equal, thus energy balance will be reached.
At this point and for each $1\leq i\leq N$ slice $ S_i $ treats a total of $ F_i + J_i $ messages, of which $ F_i $ are being slided 
and $J_i$ are being ejected. The output of the algorithm representing the randomized strategy the network should apply
is therefore going to be the following ordered sequence: $\set{ \frac{ F_i }{ F_i + J_i } }_{1\leq i\leq N}$
\subsection{Special cases}
In this section, we lift the restrictions following from the assumption that all $\epsilon$'s are positive
(remark \ref{positive:epsilons}), and from the assumption that equation (\ref{cond:1}) holds, starting with the former.
\subsubsection{First Case: Little Messages}
\label{first:problem}
Suppose that while executing the algorithm from previous section, equation (\ref{cond:1}) does not hold for some $i$.
What this means is that for some $i$, while treating slice $S_i$, even if  all the $ g_i $ generated messages
are ejected to the sink,
$S_i$ will not spent as much average energy per sensor as slice $ S_{ i-1 } $ does;
informally: \emph{there are not enough generated messages at slice $S_i$} (thus the title of this subsection). 
So what can we do?
In essence, the solution is to get slices previous to $ S_i $ (i.e. $S_{ i+1 }$, $ S_{ i+2 }$, etc...)
to forward some of there generated messages towards $ S_i $, so that $ S_i $ can ``catch-up'' with $ S_{ i - 1 }$.
Practically, this is done by recursively using the algorithm described erstwhile.
To bring this idea to effect, we \emph{go down} one-level in the recursion. This brings the need to stack
some values, which we will be able to unstack when \emph{coming up} one level in the recursion, and to compute
some new values.
The following enumeration stepwise explains how to implement this scheme.
\begin{enumerate}
  \item
  We start by stacking previous values and computing new ones.
  \begin{enumerate}
  \item\label{compute:new:start} 
    Stack the current value of a variable $start$, which remembers at what position the last recursion
    started. If this is the first level of recursion,
    we stack the value $start = 1$.
  %\item
    The new value of $start$ is set to $start = i $, which is the position of the slice which is trying to catch-up.
  \item Stack the current value of a variable $maxNrg$. If this is the first level of recursion,
    we stack the value $maxNrg = \infty$. The new value of $maxNrg$ is
    set to
  %\item Now, compute a new current value for $maxNrg$: 
    \[maxNrg = \frac{1}{ b_{ start - 1 } }\left( F_{ start - 1 } + J_{ start - 1 }d_{ start - 1 }^2 \right)\]
    which is the amount of energy to be spent by $S_{start}$ in order to ``catch-up'' with $S_{start-1}$.
  \item\label{compute:new:epsilons}
    Stack the value of the previous $ \epsilon $'s.
  %\item \label{catch:up} 
    Set new current values for $\epsilon_j$'s for $ start \leq j \leq N$:
    \begin{itemize}
      \item
	$ \epsilon_{start} $ is set to $ 1 $ (slice $S_{start}$ ejects every sliding messages it sees, in order to try catching-up with 
	$S_{ start-1 }$).
      \item \label{recompute:epsilons}
	The other $\epsilon_k$'s with $start < k \leq N $ are computed using equation (\ref{rec:epsilon:fine}).
    \end{itemize}
  \end{enumerate}
  \item 
    Next, we go down one level in the recursion, which essentially means
    redoing the algorithm from section \ref{algorithm}, but using the above newly computed $\epsilon$'s,
    and considering the possibility of either going further down one recursion level,
    or on the contrary coming back up one recursion level.
    Before describing in more details what our algorithm does whilst
    going up/down recursion levels (as described above), let us make
    the following remark.
    \begin{remark}
      While in the lower level of recursion and while treating slices $S_{start+1},S_{start+2},$ etc..., three cases may occur
      (for the sake of comprehensiveness, suppose we are treating slice
      $ S_k $ for some $ k  > 0 $):
      \begin{itemize}
	\item Slice $S_{start}$ may be able to catch-up (using
	  messages previously slided from slices $S_{start+1}$ to
	  $S_{k-1}$ and the newly messages slided from $S_k$),
	  i.e. sufficiently messages will have been slided from
	  its previous slices. In this case, we shall have to go back up one recursion level.
	\item $S_k$ may in turn not be able to spend as much energy as
	  $ S_{ k - 1 } $, in this case, we need to further go down another level.
	\item Finally, neither of the two cases above may happen, slices $S_k, S_{k-1}, \ldots, S_{start}$ are spending the same amount of
	  average energy per sensor (but less than $S_{start-1}$), and we start treating the next slice, $S_{k+1}$
      \end{itemize}
    \end{remark}
    We now describe in more details the way the algorithm from section \ref{algorithm}
    should be adapted in order to implement the above remark.
    Suppose we are in the first recursion level. We are treating slice $ S_k $ for some $k>start$.
    Slice $S_{start}$ is trying to catch-up with slice $S_{start-1}$, and all the slices from $S_{start}$ to $S_{k-1}$
    are spending the same average energy per sensor (which we can suppose to be true by induction).
      \item 
	\label{start:algorithm}
	\emph{If} there are not enough messages for $S_k$ to spend as much energy as $S_{k-1}$,
	i.e. if $ \frac{1}{b_k}G_kd_k^2 < \frac{1}{b_{k-1}}\left( F_{k-1} + J_{k-1}d_{k-1}^2\right)$
	we just do our best: we eject $J_k = G_k $ messages, set $G_k = 0$, and further go down one recursion
	level. Now $S_k$ will have to try to catch-up with $S_{k-1}$.
    \item
	\label{first:step}
	\label{not:enough:messages:again}
	\emph{Else},
	we eject $J_{k} =  \frac{b_k}{d_k^2b_{k-1}}\left( F_{k-1} + J_{k-1}d_{k-1}^2\right) $ messages from $ S_{k} $ 
	and set $ G_k = G_k - J_k  $.
	$ S_{ k }$ now has the same mean energy consumption
	per sensor as $ S_{k-1} $, but there are still $G_k$ messages to be treated.
	\begin{remark}
	  \label{how:to:slide:wisely}
	Essentially, we now want to slide the remaining $G_k$ messages \emph{along the network}, from $S_k$ to $S_{start}$,
	but with some precaution: 
	\begin{itemize}
	  \item First of all, we still have to take into the account the $\epsilon$'s and
	    eject ``$\epsilon$ fractions'' of the messages sliding along the network from
	    $S_k$ to $S_{start}$.
	    This was explained in detail in equations (\ref{first:of:many}) to (\ref{last:of:many}),
	    and we therefore we consider from now on this to be
	    implicit whenever we talk of \emph{sliding messages along
	    the network}.
	  \item
	    If we have enough remaining messages, i.e. if $G_k$ is still sufficiently
	    large, we will be able to let slice $S_{start}$ (and slices $S_k$ to $S_{start+1}$) catch up with $S_{start-1}$.
	    If this is so, we want to slide \emph{just enough}
	    messages so as to catch-up, 
	    then go back up one recursion level\footnote{
	      This is to ensure that we can unstack the appropriate
	      $\epsilon$'s while coming-up a recursion level.
	      E.g. there is no more reason to use the $\epsilon$'s 
	      where $\epsilon_{start}$ was forced to the value 
	      $\epsilon=1$ if $S_{start}$ has managed to catch-up with $S_{start-1}$.},
	    and only then shall we slide the eventually remaining
	    messages along the network.
	    \end{itemize}
	\end{remark}
	Here is how we propose to implement the above remark.
      \item
	\label{slide:the:rest}
	Set $ \Delta_{nrg} = maxNrg - \frac{1}{b_k}\left( F_k + J_k d_k^2 \right)$
	and $ msgToGoUp = \frac{ b_k \Delta_{ nrg }}{ \epsilon_k d_k^2 + \left( 1 - \epsilon_k \right)  }$.
	Let $\Phi = \max \set{ G_{ k },\ msgToGoUp } $.
	From the remaining $ G_k $ messages, we further slide and eject respectively
	$ F = \left( 1 - \epsilon_{ k } \right) \cdot\Phi$ and $ J =\epsilon_k \cdot\Phi $ messages\footnote{This 
	  ensures that slices $S_k$ to $S_{start}$ still have the same mean energy consumption per Sensor
	  and that this is bounded by the mean energy consumption per sensor of slice $S_{start-1}$
	}. Finally, we need to make the following adjustments:
	$F_k = F_k + F $, $J_k = J_{start} + J$ and $ G = G - F - J $.
	\footnote{The $F_j$'s and $J_j$'s for $start \leq j < k$ also have to be adjusted,
	acknowledging the fact that $F$ new messages are slided along the network from $S_k$, and using the 
	$\epsilon$'s, but as explained in the first point of remark \ref{how:to:slide:wisely}, we consider this to be implicit.}
      \item
	\emph{If} there are enough messages for $ S_{start} $ to
	catch-up with $S_{start-1}$
	we can go back up one recursion level:
	\begin{itemize}
	\item 
	  Unstack the previous $\epsilon$'s.
	\item
	  Unstack the previous value of $maxNrg$.
	\item
	  Unstack the previous value of $start$.
	\end{itemize}
	\item
	  \begin{itemize}
	    \item
	      \emph{If} there are no more messages to treat for slice $S_k$, i.e. if $G_k=0$,
	      we can start to treat the next slice, $S_{k+1}$.
	      This means jumping to point \ref{start:algorithm} above, but this time with $k = k+1$.
	    \item
	      \emph{Else} we need to treat the remaining $G_k$ messages.
	      This is done by jumping to point \ref{slide:the:rest} here above.
	  \end{itemize}
\end{enumerate}
What happens in the end? If the algorithms returns from all the recursive calls to the main algorithm,
it is easily seen that energy balanced is reached. Otherwise, we have a solution which
is decreasing in average energy per sensor (from slice $S_1$ towards $ S_N $), and which is ``locally'' energy balanced,
for example, we could have:
\begin{equation}
  \frac{1}{b_1}E_1 = \frac{1}{b_2}E_2 = \frac{1}{b_3}E_3 > \frac{1}{b_4}E_4 > \frac{1}{b_5}E_5 = \frac{1}{b_6}E_6\geq\ldots
\end{equation}
Although not reaching energy balance, we shall prove this solution
is optimal.
The one important thing to observe is the following:
If a recursion starts at slice $S_i$, either one of the two cases happens:
\begin{itemize}
  \item
    The algorithm returns from this recursive call and the solution is locally energy balanced:
    $\frac{1}{b_i}E_i = \frac{1}{b_{i-1}}E_{i-1}$
    \item
      The algorithm does not return from this recursive call and
      the solution is \emph{not} energy balanced:
      %\begin{equation}
	%\label{catch:up:1}
	$\frac{1}{b_{i}}E_{i} < \frac{1}{b_{i-1}}E_{i-1}$.
	%\end{equation}
      Furthermore, since $S_i$ was trying to ``catch-up'' with $S_{i-1}$ and that we set $\epsilon_i=1$ 
      (point (\ref{compute:new:epsilons}) of the algorithm), it holds that $F_i = 0 $, and thus that 
      \begin{equation} 
	\label{catch:up:2}
	p_i = 0
      \end{equation}
\end{itemize}
In section \ref{analysis}, we use equation (\ref{catch:up:2}) to show that this solution is always optimal.
\subsubsection{Second Case: Little Battery}
\label{second:problem}
The second problem which may occur is when the assumption that all $\epsilon$'s are positive, 
(i.e. the assumption from remark \ref{positive:epsilons} does not hold).
From equation (\ref{rec:epsilon:fine}), we can see that this occurs only if some of the slices have little $b_i$'s
(and thus the title of this subsection).
Let us first analyse what it means for an $\epsilon$, say $\epsilon_i$ to be \emph{negative}.
Suppose slice $ i $ has, so far, $j_i$ ejected messages and $f_i$ forwarded messages.
When it receives $k$ sliding message from $ S_{i+1}$, it should eject an $\epsilon_i$ fraction to the sink,
and pass-on the $1-\epsilon_i$ rest to the next slice. After this, there are
%\begin{itemize}
  %\item 
  $j_i + k \epsilon_i $ ejected messages and
  %\item 
 $f_i + k \left(1-\epsilon_i\right) $ slided messages.
%\end{itemize}
The fact that the $\epsilon$'s satisfy equation (\ref{rec:epsilon:fine}) ensures that energy balance is conserved
(at least locally if we are already into a recursive call as described in section \ref{first:problem}).
So what is the problem? The problem is that since $\epsilon_i$ is negative, $j_i$ becomes negative if
\begin{equation}\label{too:many:slided}
k\cdot\epsilon_i + j_i < 0 \Leftrightarrow k > j_i/\epsilon_i
\end{equation}
and thus the solution is not physical (a negative amount of messages can not be ejected from $ S_i $).
The fix to this problem has some similarity with the previous one.
Whenever a slice (say the $i$th) is about to slide $k$ messages along the network,
it should ensure that no slice will find itself in a non physical state afterwards
by bounding the number of messages it allows itself to slide along the network.
Suppose a slice $S_k$ wants to slide messages along the network, for some fixed $k$.
We call $maxSlide$ the maximum number of messages $S_k$ may slide along the network
without putting any of its following slices in a non-physical state.
In order to compute $maxSlide$, we should remember what happens when $k$ message are slided
along the network by slice $S_l$: part of it is ejected by each of the slices sliding the message, according
to the $\epsilon$'s, ans therefore only a $ k_i = k\cdot\Pi_{j=i+1}^l \left( 1 - \epsilon_i \right) $ fraction of the  $k$ initial 
messages reaches slice $S_i$. $maxSlide$ is defined as the maximum value $k$ such that 
$ k_i \epsilon_i + j_i \geq 0 $ for $ 1\leq i \leq l$,
or equivalently, the maximum value such that
$ k_i \leq \frac{ j_i }{ |\epsilon_i| } $
for every $ 1 \leq i \leq l$ such that $\epsilon_i < 0 $,
and it can be computed by the following procedure.\\

%\begin{pseudocode}[display]{computeMaxSlide}{i, start}
\begin{pseudocode}[ruled]{computeMaxSlide}{i}
  \textbf{Input: }\textrm{$i$, a slice number.}\\
  \textbf{Output: }\textrm{$max$, the max number of messages $S_i$ can slide.}\\
  \GLOBAL \epsilon[\ ] \COMMENT{This is an array storing the values of the $\epsilon$'s}\\
  \LOCAL F \GETS 1\\
  \LOCAL ejected[\ ]\\
  \FOR \left( k \GETS i; k \geq 1; k \GETS k - 1 \right) \DO
  \BEGIN
    ejected[ k ] \GETS F * \epsilon[ k ]\\
    F \GETS F * \left( 1 - \epsilon[ k ] \right)\\
  \END\\
  \LOCAL max \GETS \infty \\
  \FOR \left( k \GETS i;\ k \geq 1;\ k \GETS k - 1\right) \DO
  \BEGIN
    \LOCAL j \GETS ejected[ k ]\\
    \IF j < 0 \THEN \DO
    \BEGIN
      j \GETS -j\\
      \IF max = \infty \THEN max \GETS j[ k ] / j
      \ELSE \DO
      \BEGIN
	\LOCAL tmp\_max \GETS j[ k ] / j\\
 	\IF tmp\_max < max \THEN max \GETS tmp\_max
\END
\END
\END\\
\RETURN{max}
\end{pseudocode}\\
Once we have computed $maxSlide$, we can decide what to do when slice $S_i$ wants to slide $k$ messages.
If $ k \leq maxSlide $, then we can simply slide the $ k $ messages, but if $ k > maxSlide $,
we have to be more careful. First, we can partially fulfill the aspiration of $S_i$
by allowing it to slide $maxSlide$ messages.
At this stage, $S_i$ still wants to slide $k -maxSlide$ messages and  
one of the previous following slices (say $S_k$) has a negative $\epsilon_k$
and $j_k = 0 $. If anymore messages are slided, $S_k$ will be in a ``non-physical'' state.
So what we do is that we recompute all $\epsilon_k$'s for $1 \leq k \leq j$,\footnote{
  If we are inside a recursion of the type described in section \ref{first:problem},
  we do not recompute all $ \epsilon_k $'s for $ 1\leq k \leq N $, but we rather set $ \epsilon_{\textsl{start}} = 1 $
and recompute all the $\epsilon_i$'s for $\textsl{start} < i \leq N $, where \textsl{start} is the place where the last recursion took place
(c.f. point (\ref{compute:new:start}) of the enumeration on section \ref{first:problem}).
} 
but whenever an $ \epsilon_k < 0 $ and $ j_k = 0 $, we force $ \epsilon_k $ to $0$.
What does this do? First of all, it forces the solution to be physical.
Second, it breaks the relation from equation (\ref{rec:epsilon:fine}), since for some $k$'s
$\epsilon_k$ is forced to $0$. The fact of breaking this relation prevents slices from ejecting
a \emph{negative} amount of messages (and thus in some sense save some energy), 
when this would lead them to be in a non-physical state.
Thus slice $S_k$ will spend more energy than the (locally) energy balanced solution would require,
and on the other hand, slices following $S_i$ (that is $S_{i-1}$ to $S_1$) will spend less since the negative amount
of messages which have been prevented from being ejected where supposed to be slided along the network.
We are therefore in the presence of a ``local peak'', in the sens that: 
\begin{equation} 
  \frac{1}{b_k}E_k > \frac{1}{b_{k-1}}E_{k-1}
\end{equation}
It should be observed that for the rest, energy balance is conserved (at least locally), and furthermore
whenever such a ``peak'' appears at $S_k$, it holds that:
\begin{equation}
  \label{observation2}
  p_k = 1
\end{equation}
which is an important fact we shall use to prove that the solution obtained is optimal.
\section{Analysis}
\label{analysis}
In this section, we prove that our algorithm produces an optimal solution, in the sense that it
maximizes the \emph{lifespan} (c.f. definition \ref{df:lifespan}).
\begin{convention}
  In this section, we consider a fixed sensor network of size $N$,
  with fixed event distribution $\set{g_i}_{1\leq i \leq N}$ and fixed battery
  $\set{b_i}_{1\leq i \leq N}$. 
  A \emph{configuration} $C$ of the network is the choice of a
  sliding probability assignment $\set{p_i}_{1\leq i\leq N}$ for each slice.
  If $C$ and $\tilde C$ are two configurations, we use the letters $ f_i $ and $\tilde f_i $ to
  denote the slided messages under the configuration $ C $ and $\tilde C $ respectively.
  We do the same for the other parameter: $j_i$'s, $p_i$'s, $\epsilon_i$'s and $E_i$'s.
\end{convention}
\begin{lemma}[No Win-Win modification]
  \label{no:win:win}
  No configuration is strictly cheaper in terms of energy than another configuration:
  if $C$ and $\tilde C$ are two configurations, then there exists an $i$ such that:
  \[
  \frac{1}{b_i}E_i \geq \frac{1}{b_i}\tilde E_i
  \]
\end{lemma}
\begin{proof}
Suppose (absurd) this is not true. Therefore there exists two configurations $C$ and $\tilde C$
such that 
\begin{equation}
  \forall i\; \frac{1}{b_i}E_i \leq \frac{1}{b_i}\tilde E_i \label{absurd:hypothesis1} 
\end{equation}
and for at least one of the $i$'s
\begin{equation} 
  \label{absurd:hypothesis2} 
  \frac{1}{b_i}E_i < \frac{1}{b_i}\tilde E_i
\end{equation}
We now define the following configurations $C_0 = \tilde C $ and $ C_N = C$,
and more generally, $ C_i $ is the configuration where the $ i $ last $ p_i $'s 
(i.e. $p_N$, $p_{N-1}$, \ldots $p_{N-i+1}$) are the same as the $p_i$'s from $C$, where as the $N-i$ first $p_i$'s are
the same as the $p_i$'s from $\tilde C$.
\newcommand\Prefix[3]{\vphantom{#3}#1#2#3}
%  usage:
%  \Prefix^{A}{B} and \Prefix_{A}{B}
Then for each $1\leq i\leq N$, if we use $\Prefix^{i}{E}$ and $\Prefix^{i}{b}$ to design the energy and battery of configuration
$C_i$, the following can be seen to hold:\\
\begin{eqnarray}
  \frac{1}{\Prefix^{i}{b_k}}{\Prefix^{i}{E_k}} & = \frac{1}{b_k}{E_k} & \forall N \geq k > N - i\label{easy:equation}\\
  \frac{1}{\Prefix^{i}{b_i}}{\Prefix^{i}{E_i}} & \geq \frac{1}{b_i} {E_i}& \label{above:contradiction}
\end{eqnarray}
Where equation (\ref{easy:equation}) follows from the definition of $C_i$ and where
equation (\ref{above:contradiction}) follows from the easy observation that
$\frac{1}{\Prefix^{i}{b_i}}{\Prefix^{i}{E_i}} \geq \frac{1}{{\tilde b_i}}{{\tilde E_i}} $
combined with equation (\ref{absurd:hypothesis1}).
Next, let $k = \max\set{ i\;|\;\frac{ 1 }{ b_i } E_i < \frac{ 1 }{ \tilde b_i } \tilde E_i }_{ 1 \leq i \leq N } $,
which exists by (\ref{absurd:hypothesis2}).
Then for every $i \leq k$ the inequality in (\ref{above:contradiction}) becomes strict.
In particular, for $i=0$ (and using the fact that $C_0 = C$) it becomes   
$\frac{1}{ b_0} { E_0} > \frac{1}{b_0}{E_0}$,
which is the contradiction we need.
\end{proof}
The reason we give this lemma the \emph{no win-win modification} name is that a principle can be derived from it,
the \emph{no win-win modification principle}, which is the following: \emph{if you have a configuration and you
modify it to save energy in some parts of the network, then necessarily you will spend more energy in another part
of the network}.

In \cite{nrj:balance:jose}, the authors point out that looking at the numerical solutions, one observes that
an energy-balanced solution mostly uses single-hop data propagation, and only with little probability
propagates data directly to the sink. The authors then suggest that this is an important finding
implying that the energy-balanced solution is also energy efficient, since it only rarely uses the costly single-hop
direct ejection of messages to the sink. Our previous lemma enables to easily formalizing this intuition:
\begin{corollary}\label{balance:egal:optimal}
  Any energy-balanced solution is optimal in terms of lifespan:
  If $C$ is an energy-balanced configuration (i.e. $\forall i\;\frac{E_i}{b_i}=\frac{E_{i+1}}{b_{i+1}}$),
  then for every $\tilde C$ other configuration, we have the following inequality, with equality if and only if $ C = \tilde C $:
  $ \max\set{\frac{E_i}{b_i}} \leq \max\set{\frac{\tilde E_i}{\tilde b_i}} $, that is, $C$ maximizes the lifespan amongst all
  possible configurations.
\end{corollary}
Next we generalize corollary \ref{balance:egal:optimal}.
\begin{lemma}
  \label{lemma:table:top}
  Let $C$ be a configuration of our network. 
  Let $ max = \max\set{ \frac{ E_i }{ b_i } } $.
  Let $ k = max\set{ i\;|\;\frac{ E_i }{ b_i } = max } $\footnote{If $k=N$ the second condition hereunder is void.}.
  Let $ l = min\set{ i\;|\; \frac{ E_i }{ b_i } < max} $\footnote{If such an $l$ does not exist,
  then the second condition hereunder is void.}.
  The configuration is optimal if and only if the conjunction of the following holds:
  \begin{itemize}
    \item
      $ p_{ k + 1 } = 0 $
    \item
      $ p_{ l + 1 } = 1 $
  \end{itemize}
\end{lemma}
\begin{proof}
We only give the ideas of the proof.
First, notice that slices $S_k$ to $S_l$ forms a tabletop-like maximum of the plotting of slice position
against average energy consumption per sensor. Since $p_{k+1}=0$, nothing can be done on the left-hand side of the
tabletop to lower it. Second, since $ p_{ l+1 } = 1$, the tabletop can not rely on the slices on its right to take-on a larger part
of the message sliding towards the sink. The only solution to produce a better solution than $C$ (i.e. if $C$ was not an optimal
solution), would therefore be to modify the probabilities from $p_k$ to $p_{l+1}$, i.e. to reorganize the configuration 
``inside the tabletop''. The final point is to notice that this will break the energy balance of the tabletop, increasing
the maximum (using the principle of no win-win modifications from lemma \ref{no:win:win}).
\end{proof}
\begin{theorem}
  Our algorithm always produces an optimal solution
\end{theorem}
\begin{proof}
  We would be done if we could prove that our algorithm always produces a solution where
  the maximal is reached at a tabletop with $p_i = 0$ on the left and $p_i = 1 $ on the right,
  since this enables the call to lemma \ref{lemma:table:top}.
  To see that this is the case, the main ingredients are equations (\ref{catch:up:2}) and (\ref{observation2}).
  We leave the easy details to the reader.
\end{proof}
\bibliography{bibSensor}

\begin{thebibliography}{10}
\bibitem{antoniou} T. Antoniou, A. Boukerche, I. Chatzigiannakis, S. 
Nikoletseas and
G. Mylonas, {\em A New Energy Efficient and Fault-tolerant Protocol for 
Data Propagation in Smart Dust Networks}, in Proc. 37th Annual ACM/IEEE 
Simulation Symposium (ANSS'04), IEEE Computer Society Press, pp. 43  52, 
2004.
\bibitem{boukerche} A. Boukerche, X. Cheng, and J. Linus, {\em 
Energy-Aware Data-Centric Routing in Microsensor Networks}, in Proc. of 
ACM Modeling, Analysis and Simulation of Wireless and Mobile Systems 
(MSWiM), pp.~42-49, Sept 2003.
\bibitem{boukerche2} A. Boukerche, R. Werner, N. Pazzi and R.B. Araujo, 
{\em A Novel Fault Tolerant and Energy-Aware Based Algorithm for 
Wireless Sensor Networks}, First International Workshop on Algorithmic 
Aspects of Wireless Sensor Networks (ALGOSENSORS), Turku, Finland, 
Lecture Notes in Computer Sciences 3121, Springer-Verlag, July 2004.
\bibitem{chatz1} I. Chatzigiannakis, T. Dimitriou, S. Nikoletseas, and 
P. Spirakis,
{\em A Probabilistic Algorithm for Efficient and Robust Data Propagation 
in Smart Dust Networks},in the Proceedings of the 5th European Wireless 
Conference on Mobile and Wireless Systems beyond 3G (EW 2004), pp. 
344-350, 2004.
\bibitem{chatz2} I. Chatzigiannakis, S. Nikoletseas and P. Spirakis, 
{\em Smart Dust Protocols for Local Detection and Propagation}, in the 
Proceedings of the 2nd ACM Workshop on Principles of Mobile Computing 
(POMC), ACM Press, pp. 9-16, 2002.
\bibitem{nrj:balance:jose} C. Efthymiou, S. Nikoletseas and J. Rolim, {\em Energy 
Balanced Data Propagation in Wireless Sensor
Networks}, invited paper in the Wireless Networks (WINET, Kluwer 
Academic Publishers) Journal, Special Issue on "Best
papers of the 4th Workshop on Algorithms for Wireless, Mobile, Ad Hoc 
and Sensor Networks (WMAN 2004)", to appear in 2005.

\bibitem{leone} P. Leone, J. Rolim, {\em Towards a Dynamical Model for 
Wireless sensor Network}, First International Workshop on Algorithmic 
Aspects of Wireless Sensor Networks (ALGOSENSORS), Turku, Finland, 
Lecture Notes in Computer Sciences 3121, Springer-Verlag, July (2004).
\bibitem{nrj:balance:pierre} P. Leone, S. Nikoletseas, J. Rolim, {\em An Adaptive 
Blind Algorithm for Energy Balanced Data Propagation in Wireless Sensor 
Networks}, in the Proceedings of the First International Conference, 
DCOSS 2005, Marina del Rey, CA, USA, Lecture Notes in Computer Science 
3560, Springer Verlag, June/July 2005.

\bibitem{singh} M. Singh, V. Prasanna, {\em Energy-Optimal and 
Energy-Balanced Sorting in a Single-Hop Wireless
Sensor Network}, In Proc. First IEEE International Conference on Pervasive 
Computing and Communications - PERCOM, 2003.

\end{thebibliography}
\appendix
\section{Pseudocode of the Algorithm}
We provide hereunder a pseudode of the algorithm presented in this paper.
This pseudocode is rigourously based on a perl implementation of the algorithm
which was validated on various inputs.\\
\begin{pseudocode}[ruled]{computeOptimal}{N; g[\ ]; b[\ ]; d[\ ]}
\GLOBAL f[\ ];j[\ ];max = \infty;recLevel = 0;startPosition = 1;\epsilon[\ ]\\
\COMMENT{Following arrays used as stacks while changing recLevels}\\
\GLOBAL startPositions[\ ];maxs[\ ];epsilons[\ ]\\
\MAIN
\GLOBAL i = 0;initialG[\ ] = g[\ ];\\
    \FOR i \GETS 1 \TO N \DO \BEGIN f[ i ] = j[ i ] = 0\END\\
    \epsilon[\ ] \GETS \CALL{epsilons}{ 1 }\\
     \WHILE  i < N \DO
     \BEGIN
     i \GETS i + 1\\
     \CALL{push}{recLevels[\ ], recLevel}\\
     \IF  i= 1\\
     \COMMENT{First step}
     \THEN \DO
     \BEGIN
      	    j[ i ] \GETS g[ i ]\\
 	    g[ i ] \GETS 0
     \END\\

\ELSE\\
\COMMENT{From second step to the Nth}
\DO 
\BEGIN
\LOCAL E1 \GETS f[ i - 1 ] + j[ i - 1 ] * d[ i - 1 ]^2\\
\LOCAL ideal_j \GETS \CALL{avgNrj}{ i - 1 } * b[ i ] / d[ i ]^2\\
 	    \IF ideal_j > g[ i ]\\
	    \COMMENT{Not enough messages to stay at this level}
	     \THEN \DO \BEGIN
 		\CALL{eject}{ g[ i ] }\\
 		\CALL{downOneLevel}{} 
 	    \END\\
 	    \ELSE \\
\COMMENT{Enough messages to stay at this level}
	    \DO \BEGIN
	    \CALL{eject}{ ideal_j}\\
	    \WHILE( g[ i ] > 0 ) \DO \BEGIN
	    \LOCAL nrjDelta \GETS max - \CALL{avgNrj}{ i }\\
	    \LOCAL msgToGoUp \GETS 
			\frac{nrjDelta} 
			{ \frac{1}{b[ i ]} 
			    * ( e[ i ] * d[ i ]^2 
			    	+ ( 1 - e[ i ] ) ) }\\

	    \IF recLevel = 0 \OR g[ i ] < msgToGoUp\\
	    \COMMENT{Slide the rest}
	    \THEN\CALL{slide}{ g[ i ] }\\
	    \ELSE\\
    	    \COMMENT{Slide enough to go up one level}
	    \DO
	    \BEGIN
			\CALL{slide}{ msgToGoUp}\\
			\CALL{upOneLevel}{}
	    \END
\END
\END
\END
\END
\ENDMAIN
\end{pseudocode}\\
%\newpage
\begin{pseudocode}[display]{PROCEDURES}{ }
\PROCEDURE{avgNrj}{pos}
\RETURN{ \frac{1}{ b[ pos ] )} ( f[ pos ] + j[ pos ] * d[ pos ]^2 ) }\\
\ENDPROCEDURE
\end{pseudocode}
\begin{pseudocode}[display]{}{ }
\PROCEDURE{epsilons}{first;option}
\LOCAL \epsilon[\ ]\\
\FOR k \GETS 1 \TO first \DO \BEGIN{ \epsilon[ k ] = 1 }\END\\
\FOR k \GETS first + 1 \TO N \DO \BEGIN
  \LOCAL A \GETS \frac{ d[ k ]^2 - 1 }{ b[ k ] }\\
  \LOCAL B \GETS ( \epsilon[ k - 1 ] * ( d[ k - 1 ]^2 - 1 ) + 1 ) / b[ k - 1 ]\\
  \epsilon[ k ] \GETS \frac{ B - 1/b[ k ] }{ A + B }\\
  \IF option = "withCaution" \THEN \DO \BEGIN
 	    \IF  \epsilon[ k ] \leq 0  \AND j[ k ] = 0  \AND  k \leq i \THEN \DO \BEGIN \epsilon[ k ] = 0 \END\\
  \END\\
\END\\
\RETURN{epsilons[\ ]}\\
\ENDPROCEDURE
\end{pseudocode}
\begin{pseudocode}[display]{}{ }
\PROCEDURE{upOneLevel}{}
\LOCAL max \GETS \CALL{pop}{maxs[\ ]}\\
startPosition \GETS \CALL{pop}{startPositions[\ ]}\\
\LOCAL tmp[\ ] \GETS \CALL{pop}{epsilons[\ ]}\\
\epsilon[\ ] \GETS tmp[\ ]\\ 
recLevel \GETS recLevel - 1\\
\ENDPROCEDURE
\PROCEDURE{downOneLevel}{}
\CALL{push}{maxs[\ ], max}\ \COMMENT{store old max}\\
max \GETS \CALL{avgNrj}{ i - 1 }\\    
\CALL{push}{startPositions[\ ], startPosition}\\
startPosition \GETS i\\
\LOCAL tmpArray[\ ] = \epsilon[\ ]\\ 
\CALL{push}{epsilons[\ ], tmpArray[\ ]}\ \COMMENT{ store old epsilons}\\
e[\ ] \GETS \CALL{epsilons}{ i }\\
recLevel \GETS recLevel + 1\\
\ENDPROCEDURE
\PROCEDURE{eject}{eject}
j[ i ] \GETS eject\\
g[ i ] \GETS g[i] - j[ i ]\\
\ENDPROCEDURE
%\end{pseudocode}\\
%\newpage
%\begin{pseudocode}[display]{ }{ }
\end{pseudocode}
\begin{pseudocode}[display]{}{ }
\PROCEDURE{slide}{F}
\CALL{slideCarefull}{ F }
\ENDPROCEDURE
\PROCEDURE{slideCareless}{F}
g[ i ] \GETS g[i] - F\\
\FOR ( k \GETS i; k \geq 1; k \GETS k-1 ) \DO \BEGIN
f[ k ] \GETS f[k] + F * ( 1 - e[ k ] )\\
j[ k ] \GETS j[ k ] + F * e[ k ]\\
F \GETS F * ( 1 - e[ k ] )\\
%%     }
\END
\ENDPROCEDURE
\end{pseudocode}
\begin{pseudocode}[display]{}{ }
\PROCEDURE{slideCarefull}{F}
\LOCAL maxCarelessSlide \GETS \CALL{computeMaxSlide}{}\\
\IF maxCarelessSlide = \infty \OR F <= maxCarelessSlide \THEN \DO \BEGIN { \CALL{slideCareless}{ F } } \END
\ELSE \DO \BEGIN
 \CALL{slideCareless}{ maxCarelessSlide}\\
 F \GETS F - maxCarelessSlide\\
 \epsilon[\ ] \GETS \CALL{epsilons}{ startPosition, "withCaution"}\\
 \CALL{slide}{F}
%%     }
\END
\ENDPROCEDURE
\end{pseudocode}
\begin{pseudocode}[display]{}{ }
\textrm{ }
\PROCEDURE{computeMaxSlide}{ }
  F \GETS 1\ \COMMENT{ Simulated slide of one packet from the current pos ( i )}\\
  \LOCAL ejected[\ ]\\
  \FOR( k \GETS i; k \geq 1; k \GETS k -1 ) \DO
  \BEGIN
  ejected[ k ] \GETS F * e[ k ]\\
  F \GETS F ( 1 - e[ k ] )\\
  \END\\
  \LOCAL max \GETS \infty\\
  \FOR( k \GETS i; k \geq 1; k \GETS k - 1 ) \DO \BEGIN
  \LOCAL j \GETS ejected[ k ]\\
  \IF  j < 0 \THEN \DO \BEGIN 
  j \GETS -j;\\
  \IF( max = \infty )\THEN \DO { max \GETS  \frac{j[ k ]}{ j } }
  \ELSE \DO \BEGIN
    \LOCAL tmp_max \GETS \frac{j[ k ] }{ j }\\
    \IF tmp_max < max \THEN \DO max = tmp_max\\
\END
\END
\END\\
\RETURN{max}
\ENDPROCEDURE
\end{pseudocode}
\end{document}